\documentclass[sort&compress
    ,final            % use final for the camera ready runs
  ]
  {aipproc}

\layoutstyle{6x9}

\begin{document}

\title{Exact $f(R)$-cosmological model coming from the request of the existence of a Noether symmetry.}


\author{S. Capozziello}{
  address={Dipartimento di Scienze Fisiche, Universit\`{a} di Napoli
"Federico II" and INFN Sez. di Napoli, Compl. Univ. Monte S. Angelo, Ed.N, Via
Cinthia, I-80126 Napoli, Italy.}
}

\author{P. Martin-Moruno}{
  address={Colina de los Chopos, Instituto de Fisica
Fundamental, Consejo Superior de Investigaciones Cientificas, Serrano 121,
28006 Madrid, Spain.}
}

\author{C. Rubano}{address={Dipartimento di Scienze Fisiche, Universit\`{a} di Napoli
"Federico II" and INFN Sez. di Napoli, Compl. Univ. Monte S. Angelo, Ed.N, Via
Cinthia, I-80126 Napoli, Italy.} }

\begin{abstract}
We present an $f(R)$-cosmological model with an exact analytic solution, coming from the request of the existence of a Noether symmetry, which is able to describe a dust-dominated decelerated phase before the current accelerated phase of the universe.
\end{abstract}

\maketitle

%%%%%%%%%%%%%%%%%%%%%%%%%%%%%%%%%%%%%%%%%%%%
%% MAINMATTER
%%%%%%%%%%%%%%%%%%%%%%%%%%%%%%%%%%%%%%%%%%%%

In order to explain the large scale structure and the current acceleration of our Universe in the framework of General Relativity (GR), it is necessary to consider huge amounts of ``dark matter'' and ``dark energy'', having both an unknown nature.
Since, the validity of General Relativity on large astrophysical and cosmological scales has never been tested but only assumed \cite{will}. It is therefore conceivable that both cosmic speed up and missing matter are nothing else but signals of a breakdown of this theory.

Extended theories of gravity could match the data under the economic requirement that no exotic components have to be added, unless these are going to be found by means of fundamental experiments \cite{kleinert}. The minimal choice should take into account a generic function $f(R)$ of the Ricci scalar $R$ in the Lagrangian. Of course, a consistent theory of gravity must reproduce the low energy limit where GR has been tested.

In this participation, we want to summarize the results presented in \cite{Capozziello:2008im} where a general exact solution has been shwon. This solution is obtained by means of the so called \ \textquotedblleft Noether
Symmetry Approach\textquotedblright \cite{cimento} and
matches the two main important requirements that a cosmological solution
should achieve to agree with data: a transient Friedmann dust-like phase,
needed for structure formation, and an asymptotic accelerated behavior.

The general action of $f(R)$-theories can be expressed as follows
\begin{equation}
\mathcal{A}=\int d^{4} x\,\sqrt{-g}\,f(R)+\mathcal{A}_{m}\, ,
\end{equation}
where $f(R)$ is a generic function of the Ricci scalar $R$ and $\mathcal{A}_{m}$ is the action for a perfect fluid minimally coupled with gravity. In the metric formalism this action leads to 4th order differential
equations and GR is recovered in the particular case $f(R)=-R/16\pi G$.

In a Friedman-Robertson-Walker (FRW) space, we can consider the configuration space $\mathcal{Q}=\{a,R\}$ and the related tangent bundle $\mathcal{TQ}=\{a,\dot{a}, R, \dot{R}\}$ on which the canonical Lagrangian $\mathcal{L}
=\mathcal{L}(a,\dot{a}, R, \dot{R})$ can be defined.
The variables of this space are the scale factor $a(t)$ and the Ricci scalar $R(t)$ in the FRW metric. Since the Ricci scalar is not independent of the scale factor, one can use the method of the Lagrange multiplier to set $R(t)$ as a constraint of the dynamic. Therefore, we have
\begin{equation}
\mathcal{A}=2\pi^{2}\int dt\,a^{3} \left\{ f(R)-\lambda\left[ R+6\left(
\frac{\ddot a}a+\frac{\dot a^{2}}{a^{2}}+\frac k{a^{2}}\right) \right]
\right\}  .
\end{equation}

The variation of the action with respect to $R$ gives the value of the Lagrange multiplier, $\lambda=f_{R}$, where a subscript $R$ denotes differentiation with respect to $R$.
Taking into account this value in the action and integrating by parts we can obtain the point-like Lagrangian, which is a canonical function of two coupled fields, $R$ and $a$, both depending on time. This is
\begin{equation}
\mathcal{L}= a^{3}\,(f-f_{R}\,R)+6\,a^{2}\,f_{RR}\,\dot R\,\dot a
+6\,f_{R}\,a\,\dot a^{2}-6k\,f_{R}\,a\, .\label{eqz0}%
\end{equation}
The total energy $E_{\mathcal{L}}$, corresponding to
the $\{0,0\}$-Einstein equation, is
\begin{equation}
\label{energy}E_{\mathcal{L}}=6\,f_{RR}\,a^{2}\,\dot a\,\dot R+ 6\,f_{R}%
\,a\,\dot a^{2}- a^{3}\,(f-f_{R}\,R) +6k\,f_{R}\,a=D\, .
\end{equation}
where $D$ represents the standard amount of dust fluid as, for example,
measured today.

In order to find a solution for the equations of motion we ask for the existence of a Noether symmetry, which in general can be written in this form
\begin{equation}
X=\alpha\frac{\partial}{\partial a}+\beta\frac{\partial}{\partial R}
+\dot{\alpha}\frac{\partial}{\partial\dot{a}}+\dot{\beta}\frac{\partial
}{\partial\dot{R}},
\end{equation}
where $\alpha$ and $\beta$ are functions of the scale factor and of the Ricci scalar such that the Lie derivative of the Lagrangian is zero, i.e. $\mathcal{L}$ is
conserved and $X$ is a Noether symmetry. One possible solution for this constraint is
\begin{equation}
\label{sym}\alpha=1/a\quad;\quad\beta=-2R/a^{2}\quad;\quad f(R)=-\left\vert
R\right\vert ^{3/2};
\end{equation}
where the absolute value is needed because the convention with the Ricci scalar less than zero is used.

\begin{center}
\begin{figure}
\includegraphics[
height=1.95in,
width=4in
]{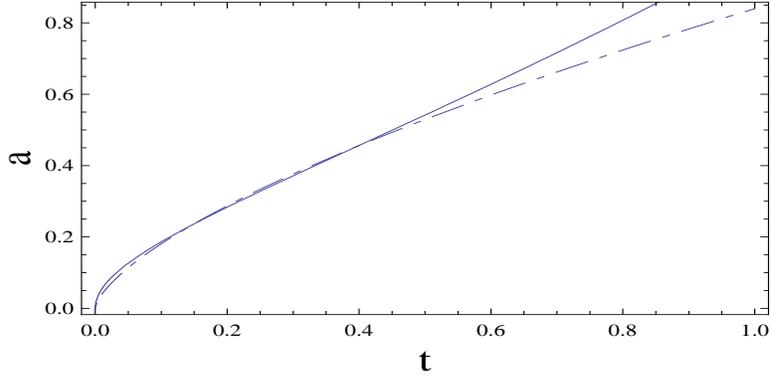}
 \caption{Scale factor versus time in standard model
(dashed) and our model\label{figura1}
(continuous).}
\end{figure}
\end{center}

Since for this $f(R)$ there is a Noether symmetry, we have an additional constant of the motion, and it must exist a change of variables $\{a,R\}\rightarrow\{u,v\}$, such that one of the new variables is cyclic. One possible change is
\begin{equation}
u=a^{2}\left\vert R\right\vert \quad;\quad v=a^{2}/2.
\end{equation}
With the new variables $u$ and $v$ it is easy to solve the equations of motion. Coming back to $a(t)$ and setting, for the sake of simplicity, $a(0)=0$, we have
\begin{equation}\label{solution}
a=\sqrt{a_{4}t^{4}+a_{3}t^{3}+a_{2}t^{2}+a_{1}t},
\end{equation}
with%
\[
a_{4}=\frac{\Sigma_{1}^{2}}{144}\quad;\quad a_{3}=\frac{\Sigma_{1}\Sigma_{0}%
}{36}\quad;\quad a_{2}=\frac{\Sigma_{0}^{2}}{24}-k\quad;\quad a_{1}%
=\frac{\Sigma_{0}^{3}}{36\Sigma_{1}}-2k\frac{\Sigma_{0}}{\Sigma_{1}}+\frac
{4D}{9\Sigma_{1}}.
\]
and $\Sigma_{0}$, $\Sigma_{1}$ and $D$ are integration constants of the equations.
It can be noted that this solution behaves as $a\propto t^{2}$ for large $t$, and as $a\propto
t^{1/2}$, for small $t$. Therefore, this solution could pass through a period during the solution approximates reasonably well a Friedmann dust-transient like $a_{f}\propto t^{2/3}$. In order to see if this transient phase is long enough to allow the structure formation, we must choose some consistent values for the integration constants.
First we fix, without lost of generality, time unities so that the current time $t_{0}=1$. This only affect to the result in the value of the Hubble parameter, since the dimensionless quantity $H_0t_0$ must have a value close to $0.93$. For simplicity, we take $H_{0}=1$. We can set $a(0)=1$, also without lost of generality, and a reasonable deceleration parameter $q_{0}=-0.4$. These considerations yield a model depending only on one
parameter. Taking $a_{4}=0.106$, the scale factor is
\begin{equation}
a=\sqrt{\frac{t}{5}[2+0.53(t-1)^{3}+t+2t^{2}]}.\label{sec2_1}
\end{equation}
If we compare the evolution of our model with the Friedmann-matter model, we obtain a very good coincidence, Fig.~\ref{figura1}. In fact the difference is close to 3\% in the red-shift interval $2\leq
z\leq4$, enough for a phase dominated by galaxies.

\begin{center}
\begin{figure}
\includegraphics[
height=1.95in,
width=4in]{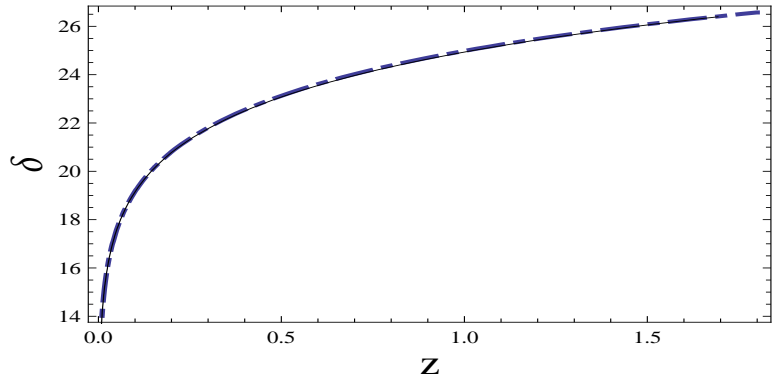}\caption{Comparison of the distance modulus $\delta$. Of our model
(continous) and $\Lambda CDMT$ (dashed).}\label{figura3}
\end{figure}
\end{center}

Now we consider the distance modulus given by the SNIa and we compare our solution with the standard $\Lambda$CDM model, as it fits data very well. Taking as reference the standard $\Lambda$CDM with a current matter density parameter $\Omega_{m,0}\simeq0.27$, we see that the coincidence is so good that it is difficult to distinguish between the two models, Fig.~\ref{figura3}.

It is interesting to pay attention to the current matter content in our model. The dimensionless parameter $\Omega_{m,0}$ must be calculated in a modify gravity theory taking into account $G_{eff}=1/\left(2f_{R}\right)$, which implies $\Omega_{m,0}\simeq0.042$.
We can see that the value for the current matter density parameter is very close to the expected for the baryonic matter in the Universe.
If we consider an observer living in a universe described by this model, who is unaware of the fact that the dynamic of his universe is described by this $f(R)$-theory, he would calculate the matter density parameter using $G_{N}$ (and not with $G_{eff}$). He would obtain $\Omega_{m,0}^{\prime}\sim0.29$, which is the value expected for all the matter content in the Universe, included the dark matter.
Therefore, in this model, it seems that taking into account dark matter could be nothing else but an assumption due to the ignorance of the physical theory behind the cosmological model.

In summary, the Noether symmetry approach allows us to obtain an analytic general solution (\ref{solution}), which interpolates between the qualitative behaviour of a Friedmann radiation-like universe, at small t, and accelerated expansion, at large t. Therefore, this solution could pass through a period during the solution approximates reasonably well a Friedmann dust-transient phase. A first attempt in the selection of the values of the parameters allows us to fulfill some observational prescription. Finallly, we would like to point out that a more accurate study and selection of the parameters is required.

%%%%%%%%%%%%%%%%%%%%%%%%%%%%%%%%%%%%%%%%%%%%%%%%
%% BACKMATTER
%%%%%%%%%%%%%%%%%%%%%%%%%%%%%%%%%%%%%%%%%%%%%%%%

\begin{theacknowledgments}
PMM gratefully acknowledges the financial support provided by the I3P
framework of CSIC and the European Social Fund.
\end{theacknowledgments}



\bibliographystyle{aipproc}   % if natbib is available


\IfFileExists{\jobname.bbl}{}
 {\typeout{}
  \typeout{******************************************}
  \typeout{** Please run "bibtex \jobname" to optain}
  \typeout{** the bibliography and then re-run LaTeX}
  \typeout{** twice to fix the references!}
  \typeout{******************************************}
  \typeout{}
 }

%%%%%%%%%%%%%%%%%%%%%%%%%%%%%%%%%%%%%%%%%%%
%% The following lines show an example how to produce a bibliography
%% without the help of the BibTeX program. This could be used instead
%% of the above.
%%%%%%%%%%%%%%%%%%%%%%%%%%%%%%%%%%%%%%%%%%%

\end{document}